\documentclass[runningheads]{cl2emult}

\usepackage{makeidx}  
\usepackage{graphicx} 
\usepackage{subeqnar} 
\usepackage{multicol} 
\usepackage{cropmark} 
\usepackage{eso}      
\makeindex            

\def\la{\mathrel{\hbox{\rlap{\hbox{\lower4pt\hbox{$\sim$}}}\hbox{$<$}}}}
\def\ga{\mathrel{\hbox{\rlap{\hbox{\lower4pt\hbox{$\sim$}}}\hbox{$>$}}}}
\def\mnras{MNRAS}
\def\apj{ApJ}
\def\aj{AJ}

\begin{document}
\title*{The 2dF QSO Redshift  Survey -- 10K@2K!}
\toctitle{The 2dF QSO Redshift  Survey -- 10K@2K!}
%
%
\titlerunning{The 2dF QSO Redshift  Survey -- 10K@2K!}
%
\author{T. Shanks\inst{1}
\and B.J. Boyle\inst{2}
\and S.M. Croom\inst{2}
\and F. Hoyle\inst{1}
\and N. Loaring\inst{3}
\and L. Miller\inst{3}
\and P.J. Outram\inst{1}
\and R.J. Smith\inst{4}}
\authorrunning{T. Shanks et al.}
%
%
\institute{University of Durham, South Road, Durham DH1 3LE, England
\and AAO, PO Box 296, Epping, NSW 2121, Australia.
\and University of Oxford, 1 Keble Road, Oxford, OX1, UK.
\and MSSSO, Private Bag, Weston Creek, ACT 2611, Australia
}

\maketitle              

\begin{abstract}
With  $\approx$10000 QSO redshifts, the 2dF QSO Redshift Survey (2QZ) is already the
biggest individual QSO survey.  The aim for the survey is  to have $\approx$25000 QSO
redshifts,  providing an order of magnitude increase in QSO clustering
statistics. We first describe the observational parameters of the 2dF QSO survey.
We then describe several highlights of the survey so far; we present new
estimates of the QSO luminosity function and the QSO correlation function. We
also present the first estimate of the QSO power spectrum from the 2QZ catalogue,
probing the form of the fluctuation power-spectrum out to the
$\approx$1000h$^{-1}$Mpc scales only previously probed by COBE. We find a power
spectrum which is steeper than the prediction of standard CDM and more consistent
with the prediction of $\Lambda$-CDM. The best-fit value for the power spectrum
shape parameter for a range of cosmologies is  $\Gamma=0.1\pm0.1$. Finally, we 
discuss how the complete QSO survey will be able to constrain the value of
$\Omega_\Lambda$ by combining results from the evolution of QSO clustering and
from a geometric test of clustering isotropy.
\end{abstract}

\section{Introduction}

The observational aim of the 2dF QSO Survey is to use the new AAT 2dF fibre-optic
coupler to obtain redshifts for 25000  B$<$20.85, 0$\la$z$\la$3 QSOs in two
$75\times5$deg$^2$ strips of sky  in the RA ranges  21h50 - 03h15 at $\delta$ =
-30$^\circ$ and 09h50 - 14h50 at $\delta$ = +00$^\circ$. The 2dF instrument
allows  spectra for  up to 400 QSO candidates to be obtained simultaneously. The
input catalogue is based on APM  UBR magnitudes for $7.5\times10^6$ stars to
B=20.85 on 30 UKST fields. The final QSO catalogue will be an order of magnitude
bigger than previous complete  QSO surveys. The prime  scientific aims of the 2dF
QSO survey  are to determine the QSO clustering power spectrum,  P(k),  in the
range of spatial scales, $0\la r\la 500h^{-1}$Mpc and to measure $\Omega_\Lambda$
 by combining results from cosmological geometric distortion and the evolution of
QSO  clustering. We also aim to cross-correlate the QSOs with 2dF Galaxy Survey 
groups to measure $\Omega_0$ via gravitational lensing (cf. \cite{cs99a}). The
2QZ survey is ideal for these clustering projects because the faint magnitude limit means
that the sky density of QSOs is high ($\approx35deg^{-2}$) which makes for a
higher count of QSO pairs in correlation analyses. Other aims include determining
the evolution of the QSO LF evolution to z=3   and placing new constraints on  
$\Omega_\Lambda$ by finding the sky density of close (6-20$''$) lensed QSO pairs.

\section{2dF QSO Redshift Survey Status}
 
We now have redshifts for $\approx$10000 QSOs measured  using 2dF itself. A
further 400 QSOs selected from the input catalogue have been observed on
different telescopes for associated projects,  including  117 bright 17$<B<$18.25
QSOs using UK Schmidt Telescope FLAIR fibre coupler, 111 radio-loud QSOs
identified in the NRAO VLA Sky Survey (NVSS) and observed at Keck and finally 30
QSOs in close pairs ($<20''$) from the ANU 2.3-m  telescope. This makes the 2dF
survey already the biggest  individual QSO survey by a factor of $\approx$10.

\begin{figure}
\centering
\includegraphics[width=\textwidth]{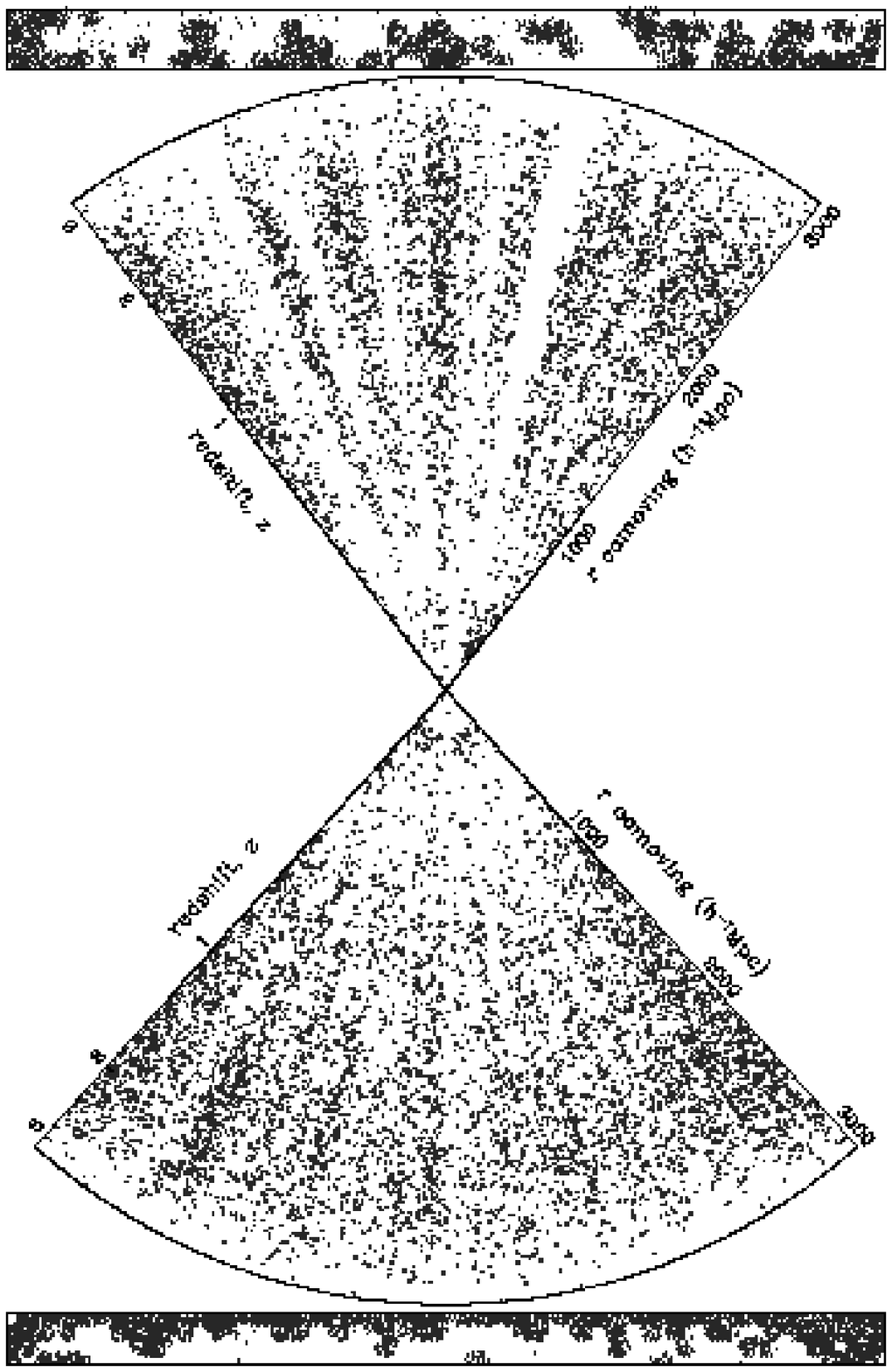}
\caption[]{The distribution of $\approx$10000  QSOs with currently measured
redshifts in the 2dF QSO Redshift Survey. The top wedge is the NGP area at
$\delta=0^{\circ}$, while the bottom wedge is the SGP region at
$\delta=-30^{\circ}$.   The rectangular strips show the survey's sky coverage up
to July 2000.}
\label{wedge} 
\end{figure}

The fraction of  our colour selected candidates which are QSOs is 54\% which
means there will be 26000 QSOs in  final survey. The QSO number count, n(B), has
been found to be  in good agreement  with previous surveys. The QSO redshift
distribution, n(z), extends to z$\approx$3 because of our multi-colour UBR
selection. In Fig. \ref{wedge} we show the current Northern and Southern redshift
wedges from the 2dF survey. The rectangles shows the sky distribution of  the
fields that have already been observed. The potential of the survey for probing
the large scale structure of the Universe is illustrated  by the scale at the edge
of the wedges reaching a comoving distance of 3000h$^{-1}$Mpc.

\begin{figure} 
\centering
\includegraphics[width=.8\textwidth]{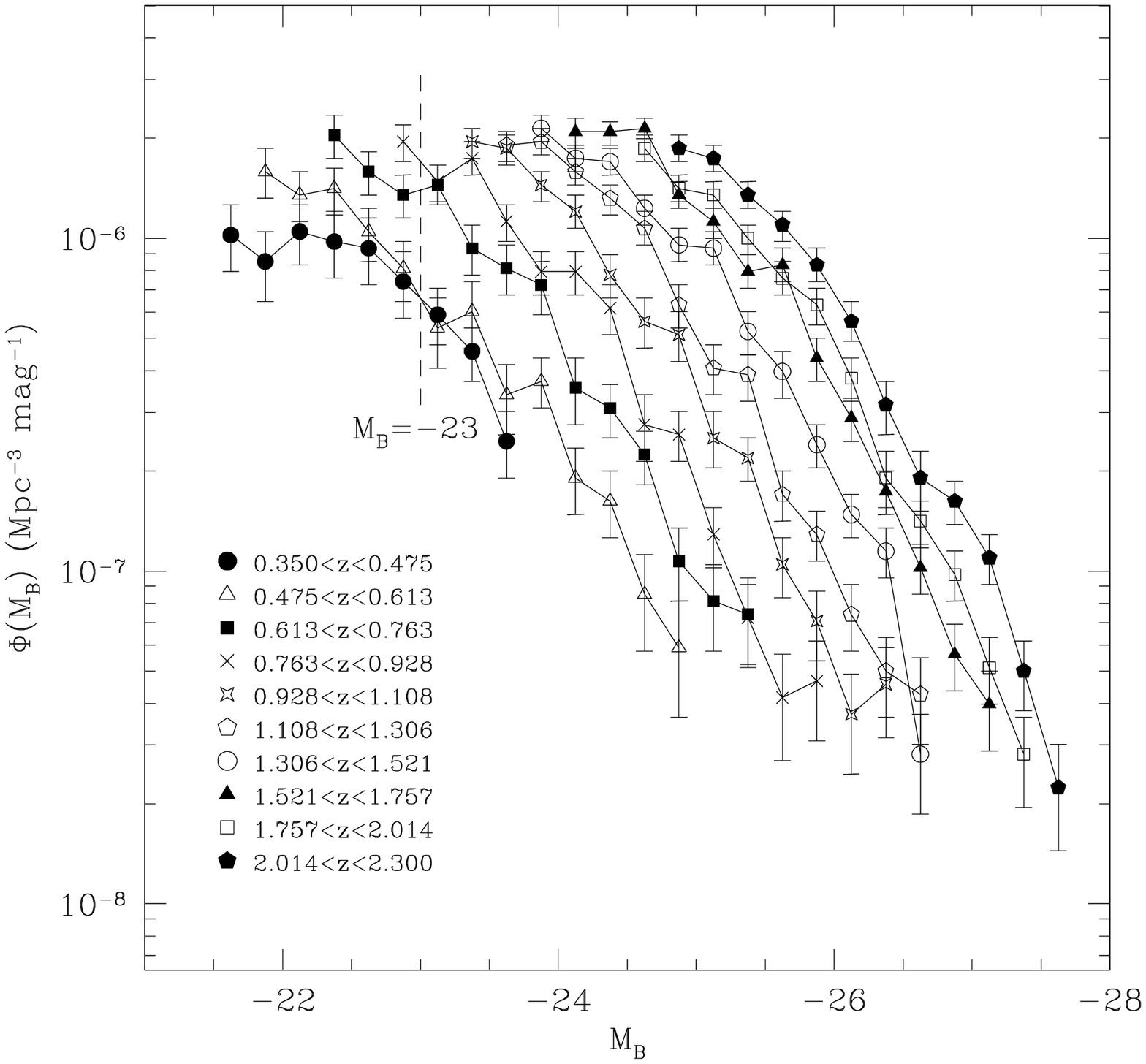}
\caption[]{The 2dF + LBQS  QSO Luminosity Function based on $\approx$6000 QSOs
for a q$_0$=0.5, H$_0$=50kms$^{-1}$Mpc$^{-1}$ cosmology. Incompleteness at
M$_B>-23$ is due to host galaxy contamination.}
\label{olf} 
\end{figure}

We have detected 8 close ($<20''$) QSO pairs. Only one is a candidate
gravitational lens; the separation is $\approx16''$ \cite{garch}.
However, the total number of lensed QSO pairs must await the
completion of the survey.

The most interesting individual QSO that has been discovered from the 2dF QSO
survey is UN J1025-0040, a  unique, post-starburst radio QSO at z=0.634,
identified in the 2dF-NVSS catalogue and followed up spectroscopically  at the
Keck \cite{broth99}. As well as broad emission lines, the spectrum
also shows strong Balmer absorption lines indicative of a post-starburst galaxy.
The starburst component of the spectrum at M$_B$=-24.7 dominates the AGN
continuum spectrum by $\approx$ 2mag. This 2dF-NVSS collaboration  has previously
also uncovered a new class of radio-loud BAL QSOs \cite{broth98} and
clearly has great potential for further exciting discoveries.

The new 2dF results for  the QSO Luminosity Function~\cite{lf}  continue to  be
consistent with Pure Luminosity Evolution models throughout the range
0.35$<$z$<$2.3~\cite{bjbcat,bjs}. The luminosity function based on $\approx$6000
QSOs of the 2dF survey  combined with $\approx$1000 LBQS QSOs~\cite{lbqs}  are
shown in Fig. \ref{olf}. The large sample size makes it possible to define the
QSO Luminosity Function in much smaller redshift bins than used previously and
the accuracy of pure luminosity evolution as a description of QSO evolution is
clear.

\section{The Hubble Volume Light Cone} 

To test our various QSO clustering statistics, we are using a Hubble Volume N-body
simulation, courtesy of  the Virgo Consortium. The assumed cosmology is
$\Lambda$-CDM with $\Omega_m$=0.3, $\Omega_\Lambda$=0.7 and $\sigma_8$=1.0. The
simulation contains 1 billion mass particles. Three  5$\times$75deg$^2$
light-cones have been output to z$\approx$4 providing 3 mock catalogues which are
the equivalent of one of our 2dF QSO catalogue strips\cite{fhthesis}.

A scale independent bias has been applied to produce a constant QSO clustering
amplitude with redshift, as seen in previous QSO surveys. The way this was done
was to calculate the space density of particles in  cubical bins of side
20h$^{-1}$Mpc and then calculate the number of QSOs in the bin according to a
threshold corresponding to a two-parameter exponential, a similar function to
that used  to bias galaxies in a $\Lambda$-CDM simulation~\cite{hat}.
QSOs are then random sampled to the same background n(z) as the observed 
catalogue. A completeness mask which accounts for the current spectroscopic
completeness can then be applied  to mimic the current 2QZ dataset.

\begin{figure} 
\centering
\includegraphics[width=.8\textwidth]{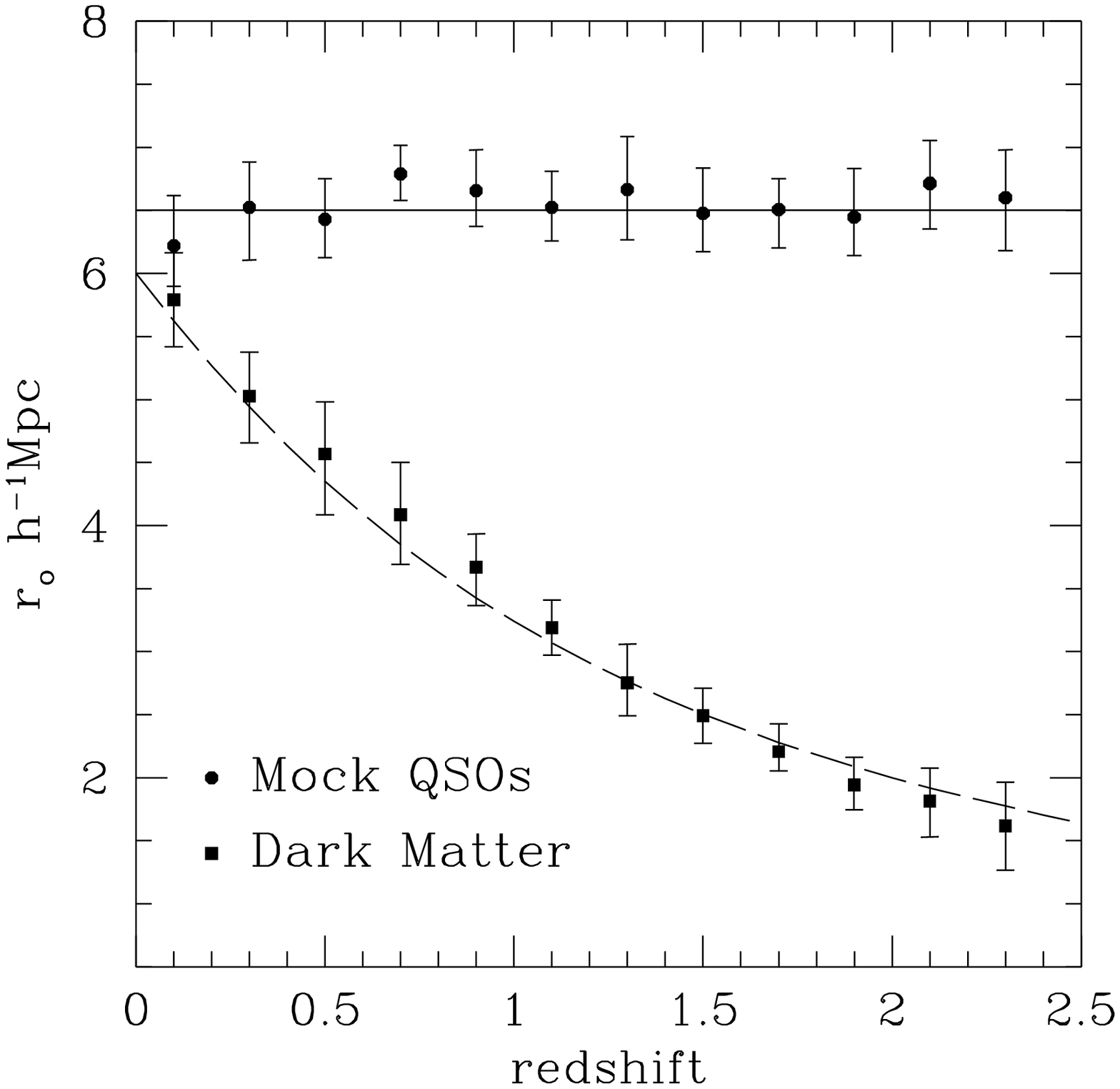}
\caption[]{Comparison of the Hubble Volume mock catalogue measured amplitude for
the mass correlation function, compared to the correlation function measured for
the biased QSOs. Both are measured in comoving coordinates. The amplitude of mass
clustering decreases rapidly with redshift whereas the clustering amplitude of
the biased QSOs remains constant with redshift by construction, to mimic previous
QSO correlation function data \cite{shanboy,cs96}. }
\label{hvamp} 
\end{figure}

Fig. \ref{hvamp} shows the comparison between the amplitude of the correlation
function of the mass and the biased, mock QSOs, as a function of redshift in the
lightcone. The QSO correlation function measured in comoving coordinates is flat
with redshift, as intended, within measurement error. The overall mock correlation
function also shows good  agreement with the 2QZ data\cite{fhthesis}.

These mock catalogues are used to test our estimators of $\xi(r)$, P(k) etc. and
to estimate their r.m.s. errors. They will also  be used  to explore the effects of
different models of bias on estimates of cosmological parameters from geometric
and redshift space clustering distortion.

It should be noted that the evidence from deep CCD imaging studies to determine
the QSO-galaxy cross-corrrelation function~\cite{eyg,smith95,cs99b,smith00} 
is that radio-quiet
QSOs, unlike radio-loud QSOs,  inhabit average galaxy clustering environments out
to z=1.5. This implies that the bias problem for radio-quiet QSOs may be no worse
than for galaxies. Indeed, there is still the possibility that radio-quiet QSOs
may even trace the distribution of optically selected L$^*$ galaxies.

\section{QSO Correlation Function}

\begin{figure} 
\centering
\includegraphics[width=.8\textwidth]{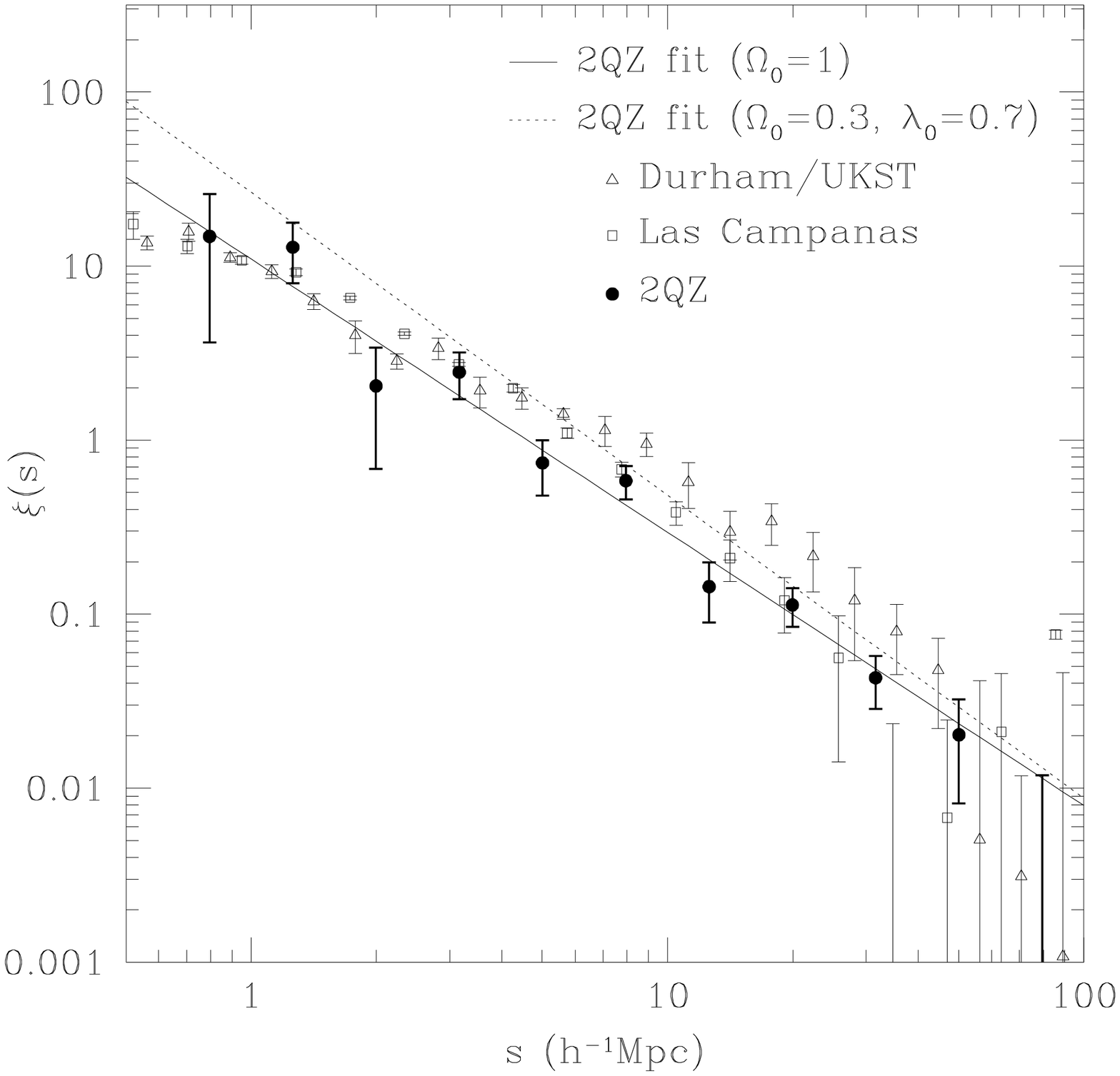}
\caption[]{The 2-point auto-correlation function for 10000 QSOs from the 2dF QSO Redshift Survey
(closed circles), measured in comoving coordinates with $\Omega_0$=1. Also plotted are the
correlation functions of local galaxies from the Durham/UKST Survey~\cite{xirat} and
the Las Campanas Survey~\cite{tucker}. The best fitting power law for two choices of
cosmology are also shown~\cite{croomxi}.} 
\label{xir} 
\end{figure}

We present a  preliminary  2dF QSO correlation function from our most complete
subset of 10000 QSOs~\cite{croomxi}. We have taken into account the current
incompleteness of the 2dF survey  as best we can; however, this process is
complicated by the fact that many observed areas still have overlapping 2dF
`tiles' as yet unobserved. We show the  correlation function in comoving
coordinates in Fig. \ref{xir}. As can be seen, the  QSO correlation function in
comoving coordinates has a power-law form with an  amplitude,
$\xi(r)=(r/r_0)^{-1.6}$ where $r_0=5.7h^{-1}$Mpc in the $\Lambda$ cosmology and
in good agreement with previous results for QSOs~\cite{cs96}. It can also be seen
that the QSO correlation functions at the average z$\approx$1.3 remain in
generally good agreement with the form of local, optically selected galaxy
correlation functions.

The correlation function amplitude shows little evolution with redshift in
$\Omega_m=1$ cosmology and appears to rise only slowly with redshift in the
$\Omega_m=0.3$, $\Omega_\Lambda=0.7$ cosmology~\cite{croomxi}. This behaviour is
consistent with evolution due to gravitational clustering  either in a low
$\Omega_0$ model or in a biased, $\Omega_0=1$ model. The correlation function is
consistently positive out to about 20h$^{-1}$Mpc at 3$\sigma$  and out to about
50h$^{-1}$Mpc at 1$\sigma$. The correlation function shows similar
results to the APM galaxy redshift survey, indicating  more power at
large scales compared to  a standard CDM model  and suggesting excess large scale
power comparable to that required in a $\Lambda$-CDM model\cite{croomxi}.

\section{First QSO Power Spectrum}

\begin{figure} 
\centering
\includegraphics[width=.8\textwidth]{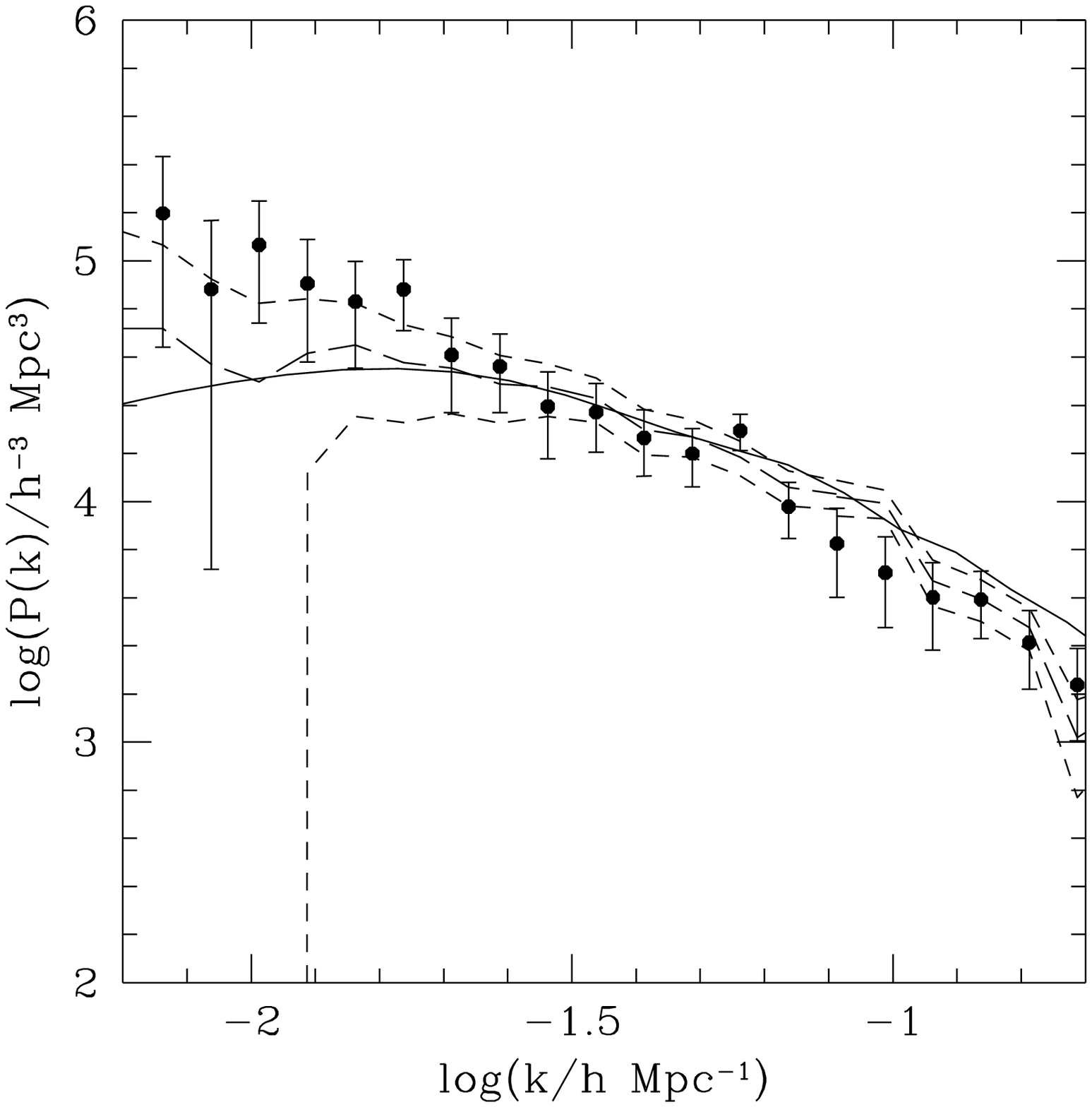}
\caption[]{ The power spectrum for $\approx$ 10000 QSOs from the 2dF QSO
redshift Survey (closed circles), measured in comoving coordinates with
$\Omega_m=0.3$, $\Omega_\Lambda=0.7$. The solid line is the input power spectrum
from the $\Lambda$-CDM mock catalogue, arbitrarily normalised. The dashed lines
are the estimated P(k) from the Hubble Volume mock catalogues with 1$\sigma$
errors.}
\label{pk} 
\end{figure}

Our estimates of P(k) in both the mock catalogues and the data are estimated by
Fast Fourier Transform at low wavenumber($log k>-1$) and direct Fourier Transform
at high wavenumber($log k<-1$) \cite{fhthesis}. The Hubble Volume mock catalogues show that even
the currently incomplete spatial window allows excellent  measurement of P(k) on
scales 50-500h$^{-1}$Mpc. Fig. \ref{pk}  shows that the differences between power
spectra measured in mock catalogues with the current completeness and window
function (dashed lines)  and the theoretical input spectrum (solid line) are
small.

We have therefore calculated the power spectrum for the $\approx$5000 2dF QSOs
in each of the N and S strips independently and they show good agreement within
the errors. We also calculated the power spectrum in a low ($0.3<z<1.4$) and a
high  ($1.4<z<2.2$) redshift bin. The low and high redshift estimates of P(k) show
only slow evolution with redshift if either the $\Lambda$-CDM or Einstein-de
Sitter cosmology is assumed, similar to the slow evolution  shown by the
correlation function.

The 2QZ power spectrum in Fig. \ref{pk} shows reasonable agreement with the Hubble Volume $\Lambda$-CDM
mock catalogue with  evidence  again of  large scale power which is in excess of what
is predicted by the standard CDM model. The best fit is $\Gamma=0.1\pm0.1$ for
both  $\Omega_m=1$  for $\Omega_m$=0.3, $\Omega_\Lambda$=0.7 cosmologies. The
power spectrum slope compares well with Durham/UKST galaxy P(k)~\cite{hoyle99} 
and with the Abell cluster, P(k)~\cite{tadros98} in both cosmologies. The
power spectrum also shows more large-scale power than the APM real-space galaxy
deprojected P(k)~\cite{be97}. 

\section{QSO Redshift Space Distortions} 

We have used the Hubble Volume mock QSO survey catalogues to test the potential
of  a  geometric  test~\cite{alcpac} for $\Omega_{\Lambda}$ in a
biased $\Lambda$-CDM model~\cite{hoyle01}. We use the correlation function split
by components of QSO separation split parallel ($\pi$) and perpendicular($\sigma$) to the 
redshift direction ($\xi_v(\sigma,\pi)$) to determine QSO redshift space  distortion.

\begin{figure} 
\centering
\includegraphics[width=.8\textwidth]{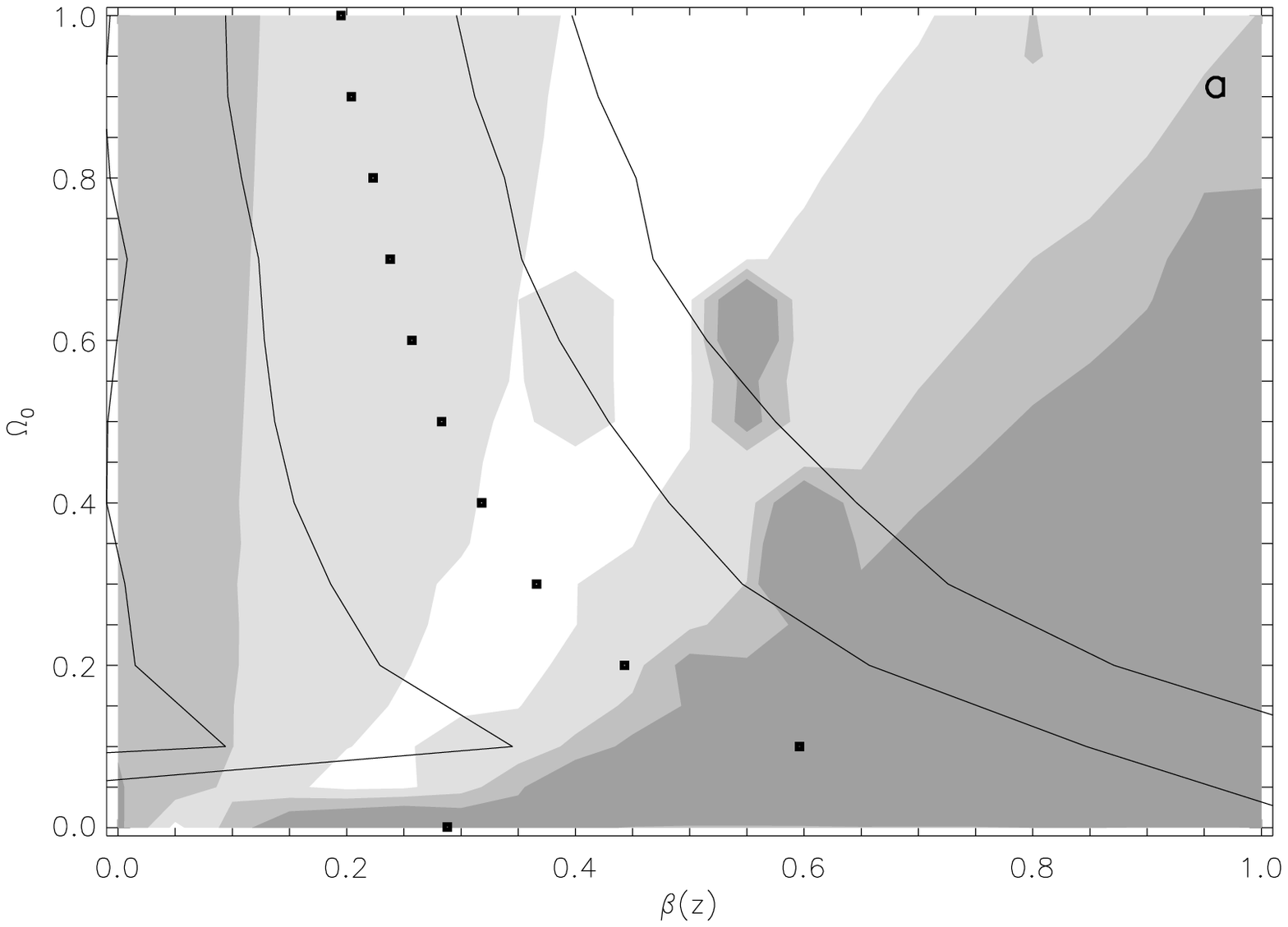}
\caption[]{Combining predicted results from measuring $\xi_v(\sigma,\pi)$ from
the Hubble Volume mock catalogue with results from the evolution of clustering.
The greyscale represents the 1,2,3$\sigma$ confidence contours in $\beta(z=1.3)$
and $\Omega_0$ from fitting geometric and linear infall distortion models to
$\xi_v(\sigma,\pi)$. The dots and lines represent the 0,1,2 $\sigma$ confidence
contours from fitting the amplitude of clustering at z=1.3 using linear
predictions for the mass evolution, normalised by observations of $\xi_{gg}$ and
$\beta_{gg}$ at z=0. The joint minimum of these confidence contours is at
$\Omega_0=0.3$ and $\beta(z=1.3)\approx0.35$ and thus the method retrieves the correct
values for the Hubble Volume mock catalogue.}
\label{xivfit} 
\end{figure}

Redshift space distortion is not only caused by peculiar velocities due to
dynamical infall but also by geometric distortion caused by a wrong assumed
cosmology~\cite{alcpac}. The initial hope was to use the latter effect to
measure $\Lambda$ but our results for the Hubble Volume suggest that this test is
degenerate in $\Omega_0$ and the linear bias factor, b, \cite{ball96}. 
This can be seen in the contour plot in Fig. \ref{xivfit} where the greyscale shows
the degeneracy in terms of $\Omega_0$ and $\beta(z=1.3)={\Omega(z=1.3)}^{0.6}/b_{qq}(z=1.3)$. 
A model which has isotropic clustering in a $\Lambda$-CDM cosmology but is then
distorted by an assumed Einstein-de Sitter geometry and linear dynamical infall
fits the similarly distorted $\Lambda$-CDM Hubble Volume data as it should but
other models with higher values of $\Omega_0$ and $\beta$ also fit.

However,  we have found that a robust test for $\Lambda$  may still be possible
by combining the $\xi_v(\sigma,\pi)$ constraint on $\Omega_0$ and $\beta$ with
the independent constraint on $\Omega_0$ and $\beta$ from the measured amplitude
of QSO clustering at our average redshift~\cite{hoyle01}. Essentially, for a
given value of $\Omega_0$, the value of $\beta_{gg}$=0.45 at z=0 which is suggested
from analyses of redshift space distortion in galaxy redshift surveys at
$z\approx0$, gives the amplitude of the mass correlation function at z=0. We then
use linear theory to calculate the correlation function of the mass at z=1.3 and
then from the observed amplitude of the QSO correlation function  we can then
calculate the value of $\beta$ at z=1.3. This gives the different
$\Omega_0$-$\beta$ degeneracy shown as the points and line contours in Fig. \ref{xivfit}.

Fig. \ref{xivfit} shows that this method to separate the $\Omega_0$-$\beta$ degeneracy
works well in the Hubble Volume mock catalogues, the overlap region of the two
sets of contours overlapping at the correct values of $\Omega_0=0.3$ and
$\beta(z=1.3)\approx0.35$. The rejection of the Einstein-de Sitter model stands at the
2-3$\sigma$ level in a survey of the size of the 2QZ. The current incompleteness
of the 2dF survey prevents the application of this analysis to the data. But an
interesting test may be in prospect in the final survey. A similar test will be 
possible at larger scales via P($k_{par},k_{perp}$), the redshift space power
spectral equivalent of $\xi_v(\sigma,\pi)$ \cite{outram01}.

\section{Future Possibilities}

We shall be also be able to use the 2dF QSO survey to make many other new
cosmological tests. For example, we will be able to make new constraints on
$\Lambda$ from counting lensed QSO pairs at both sub-arcsecond (HST) and larger
scales. We shall also be able to measure the lensing of 2dF QSOs by foreground
2dF galaxy groups and clusters via QSO-galaxy group cross-correlation to
determine $\Omega_0$ \cite{cs99a}. From obtaining higher resolution,
absorption line spectra of the brighter QSOs we shall also be able to test
whether Lyman $\alpha$ and metal absorption lines trace the same large-scale
structures as the QSOs themselves. Finally, the AAT 2dF QSO survey has shown the
exciting potential of large QSO surveys for cosmology and for probing the large
scale structure of the Universe. There is therefore strong motivation to extend
the current survey to wider fields and fainter limits using multi-colour data
from the  ESO VST and UK VISTA telescopes combined with further enhancements 
of the AAT 2dF instrument.

\section{Conclusions}

\begin{itemize}

\item The 2QZ survey now contains redshifts for $\approx$10000/26000  
B$<$20.85, z$<$3 QSOs making it already the biggest individual QSO survey by an order
of magnitude.

\item The  2QZ survey confirms the accuracy of the Pure Luminosity Evolution model f
or the evolution QSO Luminosity Function for 0$<$z$<$2.2.

\item The 2QZ survey has  already detected many individually interesting
objects, including a post-starburst QSO  and a close QSO  pair.

\item The 2QZ correlation function based on $\approx$10000 QSOs is consistent
with that for local galaxies but with more power on large scales than the
standard CDM model.

\item The first QSO power spectrum has been measured out to scales of
500h$^{-1}$Mpc. The result shows more large scale power than predicted by
standard CDM and is more consistent with the prediction of $\Lambda$-CDM.

\item 2QZ has the exciting potential to determine $\Omega_0$ and
$\Omega_\Lambda$ from redshift space  distortions, clustering evolution and gravitational lensing.

\end{itemize}

\clearpage
\addcontentsline{toc}{section}{Index}
\flushbottom
\printindex


\begin{thebibliography}{7}
%
\addcontentsline{toc}{section}{References}

\bibitem{alcpac}Alcock, C., Paczynski, B. (1979) 
An evolution free test for non-zero cosmological constant Nature, {\bf 281}, 358--359

\bibitem{be97}Baugh, C.M., Efstathiou, G. (1993) 
The Three-Dimensional Power Spectrum Measured from the APM Galaxy Survey \mnras, {\bf 265}, 145--156

\bibitem{ball96}Ballinger, W.E., Peacock, J.A., Heavens, A.F. (1996) 
Measuring the cosmological constant with redshift surveys \mnras, {\bf 282}, 877--888

\bibitem{bjbcat}Boyle, B.J., Shanks, T., Peterson, B.A. (1988)  
The evolution of optically selected QSOs. II \mnras, {\bf 235}, 935--948.

\bibitem{bjs}Boyle, B.J., Jones, L.R., Shanks, T., (1991) 
A spectroscopic survey of faint QSOs  \mnras, {\bf 251}, 482--507.

\bibitem{lf}Boyle B.J., Shanks, T., Croom, S.M., Smith, R.J., Miller, L., Loaring N., Heymans, C. (2000) 
The 2dF QSO Redshift Survey - I. The optical luminosity function of quasi-stellar objects \mnras,  {\bf 317}, 1014--1022

\bibitem{rstrans}Boyle, B.J.,  Smith, R.J.,   Croom, S.M. , Shanks, T. ,  Miller, L., Loaring, N.,  (1999) 
QSO clustering and the AAT 2dF QSO Redshift Survey Phil. Trans R. Soc. Lond. A., {\bf 357}, 185--195.

\bibitem{broth98}Brotherton, M.S., Van Breugel, W., Smith, R.J., Boyle, B.J., Shanks, T., Croom, S.M.,   Miller, L., Becker, R.H. (1998) 
Discovery of Radio-Loud Broad Absorption Line Quasars Using Ultraviolet Excess and Deep Radio Selection \apj, {\bf 505}, L7--L10.  

\bibitem{broth99}Brotherton, M.S., Van Breugel, W. Stanford, S.A., Smith, R.J., Boyle, B.J., Miller, L., Shanks, T., Croom, S.M., Filippenko, A.V., (1999) 
A Spectacular Poststarburst Quasar \apj,  {\bf 520},  L87--L90.

\bibitem{hat}Cole, S.M., Hatton, S.J., Weinberg, D.H., Frenk, C.S. (1998) 
Mock 2dF and SDSS galaxy redshift surveys \mnras, {\bf 300}, 945--966.

\bibitem{cs96}Croom, S.M., Shanks, T. (1996) 
QSO clustering - III. Clustering in the Large Bright Quasar Survey and evolution of the QSO correlation function \mnras, {\bf 281}, 893--906.

\bibitem{cs99a}Croom, S.M., Shanks, T. (1999) 
Statistical lensing of faint QSOs by galaxy clusters \mnras, {\bf 307}, L17--L21.

\bibitem{garch}Croom, S.M.,  Shanks, T.,  Boyle, B.J.,  Smith, R.J., Miller, L., Loaring, N.S.,  1999. 
In  "Evolution of Large Scale Structure: From Recombination to Garching", eds. Banday, A.J. \& Sheth,ÊR.K. E32.

\bibitem{cs99b}Croom, S.M., Shanks, T. (1999) 
Radio-quiet QSO environments - I. The correlation of QSOs and b$<$23 galaxies \mnras, {\bf 303}, 411--422.

\bibitem{croomxi}Croom, S.M.,  Shanks, T.,  Boyle, B.J.,  Smith, R.J., Miller, L., Loaring, N.S., Hoyle, F.  (2001)
The 2dF QSO redshift survey - II. Structure and evolution at high redshift \mnras, in press.


\bibitem{eyg}Ellingson, E., Yee, H.K.C., Green, R.F. (1991) 
Quasars and active galactic nuclei in rich environments. II - The evolution of radio-loud quasars \apj, {\bf 371}, 49--59.

\bibitem{lbqs}Hewett, P.C., Foltz, C.B., Chaffee, F.H. (1995) 
The large bright quasar survey. 6: Quasar catalog and survey parameters \aj, {\bf 109}, 1498--1521.

\bibitem{fhthesis}Hoyle, F. (2000)
The Structure and Scale of the Universe. PhD Thesis, University of Durham.

\bibitem{hoyle99}Hoyle, F., Baugh, C.M., Shanks, T., Ratcliffe, A. (1999) 
The Durham/UKST Galaxy Redshift Survey - VI. Power spectrum analysis of clustering \mnras, {\bf 309}, 659--671

\bibitem{hoyle01}Hoyle, F. et al. (2001) in prep.

\bibitem{outram01}Outram, P.J. et al. (2001) in prep.

\bibitem{xirat}Ratcliffe, A., Shanks, T., Parker, Q.A., Fong, R., (1998) 
The Durham/UKST Galaxy Redshift Survey - III. Large-scale structure via the two-point correlation function \mnras,  {\bf 296}, 173--190.

\bibitem{shanboy}Shanks, T., Boyle, B.J. (1994) 
QSO Clustering - I. Optical Surveys in the Redshift Range 0.3$<$Z$<$2.2 \mnras, {\bf 271}, 753--772.

\bibitem{smith95}Smith, R.J., Boyle, B.J.,  Maddox, S.J. (1995) 
The Environments of Z$<$0.3 QSOS \mnras, {\bf 277}, 270--276.

\bibitem{smith00}Smith, R.J., Boyle, B.J.,  Maddox, S.J. (2000) The environments of Z$<$0.3 QSOS \mnras, 
{\bf 313}, 252--262.

\bibitem{tadros98}Tadros, H., Efstathiou, G. (1998) 
The power spectrum of rich clusters of galaxies on large spatial scales \mnras, {\bf 296}, 995--1003

\bibitem{tucker}Tucker, D.L. et al. (1997) 
The Las Campanas Redshift Survey galaxy-galaxy autocorrelation function \mnras, {\bf 285}, L5--L9


\end{thebibliography}
\end{document}